# Structural and optical properties of beta irradiated YAlO$_3$ single crystals

M. Suganya[1], K. Ganesan[2,3,*], P.Vijayakumar[2], S. Jakathamani[3,4], Amirdha Sher Gill[1,*],

O. Annalakshmi[3,4], S.K. Srivastava[2], R.M. Sarguna[2] and S.Ganesamoorthy [2,3]

[1]*School of Science & Humanities, Sathyabama Institute of Science and Technology, Chennai, Tamil Nadu, India*
[2]*Materials Science Group, Indira Gandhi Centre for Atomic Research, Kalpakkam, Tamil Nadu, India*
[3]*Homi Bhabha National Institute, IGCAR, Kalpakkam, Tamil Nadu, India*
[4]*Health, Safety & Environment Group, Indira Gandhi Centre for Atomic Research, Kalpakkam, Tamil Nadu, India*
*\*Corresponding authors: kganesan@igcar.gov.in; amirdhashergill@gmail.com*

## ABSTRACT

We report on the growth, structural and optical properties of YAlO$_3$ single crystals grown by optical floating zone technique. Powder X-ray diffraction and Raman spectroscopic studies confirm the phase purity of the crystals. Raman analysis reveals that the intensity and line-width of Raman bands increase significantly with beta irradiation indicating the formation of structural defects in YAlO$_3$ lattice. The optical properties are studied through UV-visible absorption, and photoluminescence emission and excitation spectroscopies under pre- and post- beta irradiation. The optical studies indicate the presence of *Sm* and *Cr* impurities by exhibiting characteristic emission lines in the orange red region. Further, a systematic study on the thermoluminescence (TL) characteristics of the crystal is also carried out at different doses of beta irradiation. The crystals exhibit a prominent TL glow peak at 239 ºC for less than 5 Gy doses while a weak second glow peak evolves at higher doses. Also, the crystals show a nearly linear dose response in the studied range from 0.1 to 10 Gy. The glow curve analysis reveals that the TL emission obeys the first order kinetics model. Based on the optical studies, the plausible mechanism for the TL glow curve is discussed in terms of the intrinsic defects and impurities that are present in the crystal.

## 1. Introduction

Rare earth (RE) or transition metal (TM) ion doped yttrium aluminum perovskites (YAP or YAlO$_3$) have attracted much attention of researchers because of its outstanding optical properties with fast scintillation, high efficiency in light yield and very stable mechanical and chemical properties [1,2]. The RE or TM dopants in YAP introduce several energy levels within the forbidden bandgap and these energy levels manifest as a characteristic optical emission or absorption under suitable external stimulation which is being effectively used as solid state lasers [3]. The optical properties of YAP crystals doped with various impurities such as *Ce*, *Mn*, *Ho*, *Tm*, *Er*, *Pr*, *Cu*, *Dy*, *Sm*, *Yb*, *Cr*, *Eu* and *Nd* are extensively studied [4–19]. These doped YAP crystals also have an excellent emission characteristics in UV/visible range when irradiated with ionizing radiation and hence, it is being used in radiation dose measurements as an active element (e.g. scintillators) or passive element such as thermoluminescence (TL) detector [1,2,8,9,11]. These unique optical characteristics also make this material as promising candidate in optical data storage, holographic recording, imaging screens for electrons and X-rays and tomography systems [1,2,10,11]. It is well known that thermally stimulated luminescence is one of the powerful techniques used in dosimetric applications to monitor radiations in environmental, personal and clinical laboratories [20]. Over the decades, several TL materials are being developed for radiation dosimetry by researchers identifying new materials that exhibit better performance at an affordable cost. The doped YAP is one of the materials with several advantages. However, the studies on TL based dosimetric detectors for ionizing radiation are limited to certain dopants such as *Mn, Ce, Eu, Er, Dy, Yb, and Sm*.

TL studies are reported on Mn doped YAP single crystals which show a prominent glow peak at around 200 °C along with multiple low intensity peaks at low and high temperature regimes. Further, TL studies on *Sm,* and *Dy* doped YAP crystals and *Eu* doped YAP nanopowders are also reported and they also exhibit multiple glow curves in the temperature range from 150 to 384 °C [17,19]. Apart from single dopant, there exist a few reports on co-doping in YAP to improve the TL sensitivity [4,5,9,12,15,19]. For example, the dopant *Si* or *Hf* (which has 4+ ionic state) co-doped with Mn$^{2+}$ helps to increase the Mn$^{2+}$/Mn$^{4+}$ ratio which eventually increases the TL emission intensity and decreases the minor glow curve intensity [4,9]. Also, Mn:YAP single crystals co-doped with Cu, Co, Ce Yb and Fe are studied for improving the TL dosimetric characteristics [11,14].

Besides, there are also a few reports available on TL characteristics of undoped YAP crystals. Although the pure YAP crystals are not expected to show any response because of the non-availability of trap centers except the radiation induced point defects to store the energy from irradiation, the practical crystals always have intrinsic defects that can act as trap centers for TL response. Zhydachevskyy et al. [7] reported that the undoped YAP also exhibits two glow peaks with maximum intensity at about 225 °C along with minor peak at ~ 140 °C upon β- irradiation. TL response is attributed to the oxygen vacancy related to F and F+ color centers and $Y_{Al}^{3+}$ antisite defects [7]. Further, the undoped single crystal irradiated with 100 MeV $Si^{7+}$ ions exhibit two completely overlapped glow curves at 232 and 328 °C which are attributed to the oxygen vacancies and other point defects, such as $Y_{Al}^{3+}$ and $Al_Y^{3+}$ antisite defects [18]. Furthermore, the effect of shallow traps in undoped YAP is also studied by low temperature TL which displays seven glow curves originating from intrinsic defects that traps holes during X-ray irradiation [17].

Despite the numerous reports available in literature, it is evident that the studies on TL response of the pure or doped YAP crystals having single glow curve is scarce for high energy irradiations. Nevertheless, the materials with single glow curve have several advantages for TL dosimetric applications. In this article, we study the effect of trace level impurities and intrinsic defects on the optical and TL characteristics of intentionally undoped YAP crystals. The grown YAP crystals show a prominent single glow curve with a linear response in the studied dose range from 0.1 to 10 Gy for beta irradiation. Further, the optical properties of the pristine and beta-irradiated crystals are examined with UV-visible absorption, Raman, photoluminescence (PL) and TL emission spectroscopic measurements. The plausible mechanism for TL glow curve is discussed in terms of intrinsic defects and other uncontrolled impurities.

## 2. Experimental methods

### 2.1. Single crystal growth

Single crystals of $YAlO_3$ were grown using four mirror optical floating zone equipment (FZ-T-4000-H-HR-I-VPO-PC). The 99.995 % purity of $Y_2O_3$ and 99.999 % purity of $Al_2O_3$ were taken in stoichiometric proportion and the mixture were ground into fine powders and then sintered at 1250 °C for 24 h. Further, this mixture was ground well and again sintered at 1350 °C for 3 h. $YAlO_3$ phase formation was confirmed from powder X-ray diffraction pattern (PXRD) of the sintered materials. Subsequently, the polycrystalline seed and feed rods were prepared using $YAlO_3$ powders and then, the rods were cold pressed at isostatic pressure of 70 MPa. Once again, the pressed rods were sintered at 1250 °C for 24 h for increasing the density. Single crystal growth was carried out in air atmosphere at a flow rate of 1 l/min and growth rate of 2 - 6 mm/h with 20 - 60 rpm of feed and seed rods at counter rotation. High growth rate (6 mm/h) lead to multiple cracks in grown crystal. The crystal growth conditions were optimized by reducing the growth rate, rotation rate and air flow which yielded crack free single crystals. The wafers were cut perpendicular to the growth direction using diamond wheel and then, they were lapped and polished before taking into measurements.

**2.2. Characterization**

The X-ray diffraction and rocking curve measurements were performed using STOE XRD instrument in Bragg-Brentano geometry under theta-theta mode with CuKα X-ray source. TL glow curve analysis was carried out using RISO TL/OSL reader model DA-20. Beta irradiation was carried out using in-built $Sr^{90}/Y^{90}$ source attached to the RISO TL/OSL reader with strength and dose rate of 1.48 GBq and 0.1 Gy/s, respectively. The UV-Vis-NIR optical absorption measurements were performed using Avantes AvaLight-DH-S-BAL in the wavelength range of 200 - 1100 nm. PL excitation and emission spectra were measured using FLS980, Edinburgh Instruments. Raman and PL emission spectra were also recorded using micro-Raman spectrometer (M/s Invia, Renishaw, UK) with 532 nm excitation.

**3. Results**

**3.1 X-ray diffraction**

The typical PXRD pattern of the powders obtained by grinding YAP single crystal is shown in Fig. 1a which matches well with the reported JCPDS data [JCPDS: 01-087-1288]. PXRD pattern analysis confirms the orthorhombic structure and also affirms the single phase of the grown YAP crystals. Fig. 1b depicts the PXRD pattern of the YAP wafer displaying first and second order diffraction peaks at 26.9 and 55.5 degrees respectively, which confirm the growth orientation to be <111>. The inset in Fig.1b represents the typical X-ray rocking curve of the YAP wafer. The full width at half maximum (FWHM) of the wafer is found to be 430 arc sec which indicates the reasonably high structural quality of the crystals.

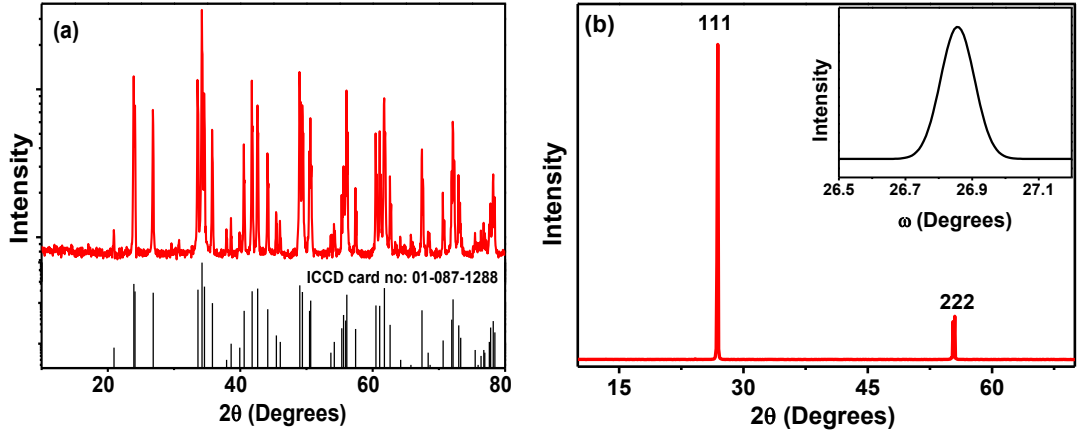

Fig 1. a) Typical X-ray diffraction pattern of the powdered YAlO$_3$ crystal. The bottom of the plot gives the JCPDS data [JCPDS: 01-087-1288] for reference b) Powder XRD pattern of YAlO$_3$ single crystal wafer. The inset shows the rocking curve plot of the crystal.

**3.2 Raman spectroscopy**

Figure 2 shows the Raman spectra of pristine and beta irradiated YAP crystals recorded under identical conditions. These spectra are subtracted for their respective background; the intensity is normalized with 429 cm$^{-1}$ band and also the spectra are overlapped with each other for an easy comparison. According to group theory considerations, the orthorhombic structure (P*nma*) of YAP has a total of 60 vibrational modes. Among them only 24 modes are Raman active with $7A_g + 5B_{1g} + 7B_{2g} + 5B_{3g}$ symmetries [21,22]. The modes corresponding to $A_g$, $B_{1g}$, $B_{2g}$, and $B_{3g}$ symmetries are marked in Fig. 2 with the label 1, 2, 3 and 4, respectively. As shown in Fig. 2, Raman spectrum of pristine YAP has 28 modes viz. 149, 157, 195, 224, 267, 270, 283, 302, 313, 327, 343, 349, 381, 404, 418, 429, 460, 473, 503, 518, 551, 563, 597, 645, 690, 735, 749, and 773 cm$^{-1}$. Of these, 20 modes labeled with 1, 2, 3 and 4 in Fig. 2 are Raman active modes which are consistent with experimental data and theoretically calculated frequencies [21,22]. Some of the remaining peaks are harmonics of the first order Raman bands and a few bands may also arise due to disorder induced IR vibrational modes. The low frequency modes below 250 cm$^{-1}$ correspond to translation motion of dodecahedron ($YO_{12}$) units without contribution from the octahedral ($AlO_6$) units while the intermediate frequencies (250-300 cm$^{-1}$) are associated with rotational motion of $YO_{12}$ in which $AlO_6$ vibrations also have an equal contribution. The high frequency modes ( > 300 cm$^{-1}$) are mainly arising from the bending modes of $AlO_6$ octahedrons [22].

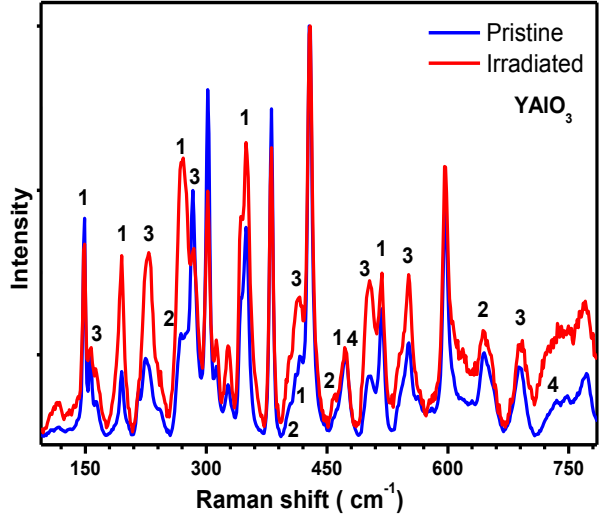

Fig. 2. Raman spectra of pristine and beta irradiated $YAlO_3$ crystals. The Raman spectra are subtracted for background and the intensity is normalized for comparison. The numbers 1, 2, 3 and 4 represent the Raman modes that arise due to $A_g$, $B_{1g}$, $B_{2g}$, and $B_{3g}$ symmetries in the $YAlO_3$ crystal.

Even after beta irradiation, all the Raman modes are present and their peak positions also remain constant. However, the intensity of the most of the Raman bands increases significantly as compared to pristine YAP crystals, as shown in Fig. 2. Especially the intensity of Raman modes at 195, 228, 270, 327, 349, 415 and 503 cm$^{-1}$ increase drastically while the intensity of the Raman modes at 149, 289, 302 and 381 cm$^{-1}$ decreases considerably. The decrease in Raman intensity is attributed to the irradiation induced structural defects / disorder and the breakage of Y-O and Al-O bonds [23]. Further, the increase in FWHM of the Raman bands also supports the increase in irradiation induced structural defects. On the other hand, the increase in Raman bands' intensity is attributed to the resonance Raman scattering phenomenon [24]. This resonance condition is satisfied because of the presence of beta irradiation induced electronic trap states with energy levels near to the Raman laser excitation energy of 2.33 eV. The energy levels of irradiation induced trap states are further discussed under optical absorption and PL spectroscopy.

### 3.3 UV-Vis spectroscopy

Figure 3 shows the UV-Vis-NIR absorption spectra of pristine and beta irradiated (2 Gy) YAP single crystals of ~ 0.15 mm thickness. These spectra are recorded at an identical conditions in the wavelength range of 220 - 750 nm. These crystals exhibit two strong absorption bands at ~ 245 and 275 nm in the UV region and also, broad and weak absorption bands at ~ 360, 480, 580, 660 and 730 nm. The oxygen vacancy is one of the common point defects in YAlO$_3$ due to high temperature crystal growth process. When the oxygen vacancy site is occupied by two (one) electrons, the color center F (F$^+$) is formed and the energy level of these defects fall within the forbidden bandgap. Hence, they absorb energy corresponding to ~ 275 and 360 nm by F and F$^+$ centers, respectively and the absorption energy levels are consistent with theoretical calculation [25]. The other absorption bands can be due to uncontrolled RE impurities that are present in the crystal. Since the 4f-5d transitions in RE ions are not parity forbidden, they have very strong absorption or emission in the UV regime.

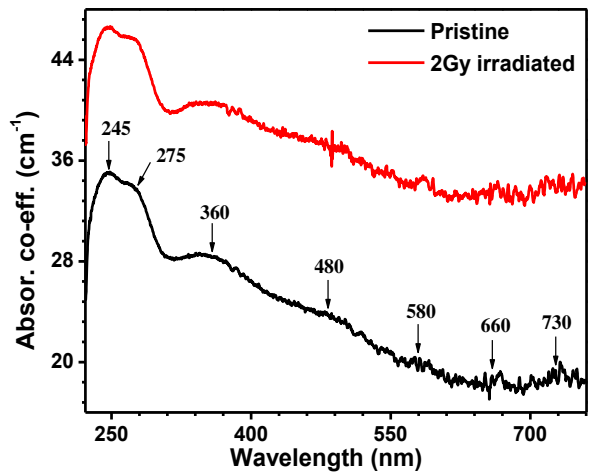

Fig. 3. Optical absorption spectra of pristine and beta irradiated YAlO$_3$ single crystal

### 3.4 Photoluminescence spectroscopy

Figure 4 shows the PL spectra of the pristine and 5 Gy beta irradiated YAP crystals with 532 nm laser excitation. The unirradiated YAP crystals do not have any emission in the wavelength range of 550 - 650 nm (curve no. 1). On the other hand, beta irradiated YAP crystal shows a strong and broad emission band in the range of 550-650 nm (curve no. 2). Further, the PL spectra of both unirradiated and irradiated crystals exhibit a sharp doublet emission peaks at 723 and 725 nm which indicates the presence of $Cr^{3+}$ impurities in YAP lattice [13]. It should also be noted here that $Cr^{3+}$ doped YAP is a potential material for optical refrigeration because of its phonon-induced anti-Stokes fluorescence that arises from zero phonon lines of $Cr^{3+}$ ions at octahedral sites [13]. In addition, several broad emission bands are also observed in the wavelength range 670 – 800 nm with notable peaks at ~ 688, 703, 715, 732 and 750 nm in pristine YAP crystals. These emission lines are also present in irradiated crystals, however, the line width of some of the peaks becomes broader and also, the intensity of a few other emission lines decreases. These observation are attributed to the beta irradiation induced structural disorder that decreases the life time of the excited carriers in the YAP lattice.

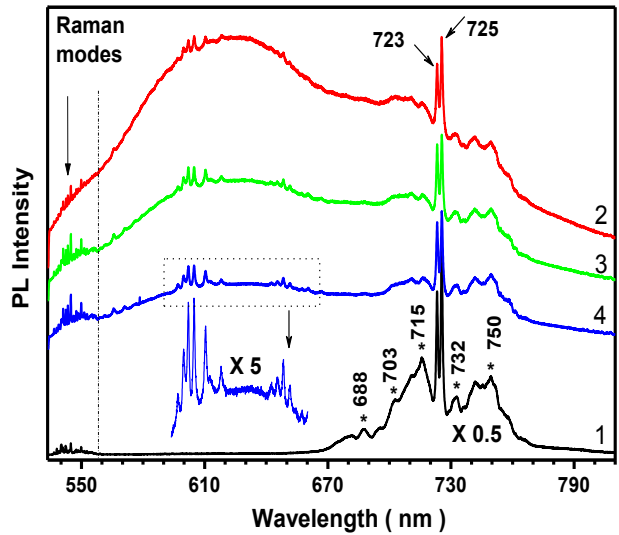

Fig. 4. Photoluminescence spectra of pristine, beta and 532 nm laser irradiated $YAlO_3$ crystals. The curves (1) and (2) represent the pristine and 5 Gy beta irradiated crystals, respectively. Curves 3 and 4, represent the PL spectra on beta irradiated crystal which is again irradiated with 532 nm laser for 30 seconds at 2 and 10 mW powers. A magnified part of curve 4 is given as inset. The spectra are recorded at identical conditions and are vertically stacked for clarity without any background correction.

Additionally, the effect of photo-bleaching behavior on PL emission is also studied by illuminating the beta irradiated YAP crystals with 532 nm laser. For this purpose, the beta irradiated YAP crystal is further irradiated with 532 nm laser for 30 seconds at a power of 2 and 10 mW. Then, the PL measurements are carried out at the laser exposed area and these spectra are shown as curves 3 and 4 in Fig. 4, respectively. As can be seen from Fig. 4, these PL intensities are lower than the laser unirradiated area (curve 2). The reduction of PL intensity of the curves 3 and 4 with respect to curve 2 indicates that some of the trapped carriers from the beta irradiation induced meta-stable states in forbidden bandgap can be released by excitation of 532 nm laser. However, the crystals retain a large PL intensity even after 10 mW indicating the stability of meta-stable energy levels against 532 nm laser irradiation. Furthermore, the PL spectra (curves 2, 3, & 4 ) consist of multiple sharp emission lines, as shown in magnified part of the curve 4 in Fig. 4 in the range of 594 – 660 nm, of the beta irradiated YAP crystal. To probe further about these emission lines, PL excitation (PLE) and emission spectra are recorded using UV light and discussed in the forthcoming paragraph.

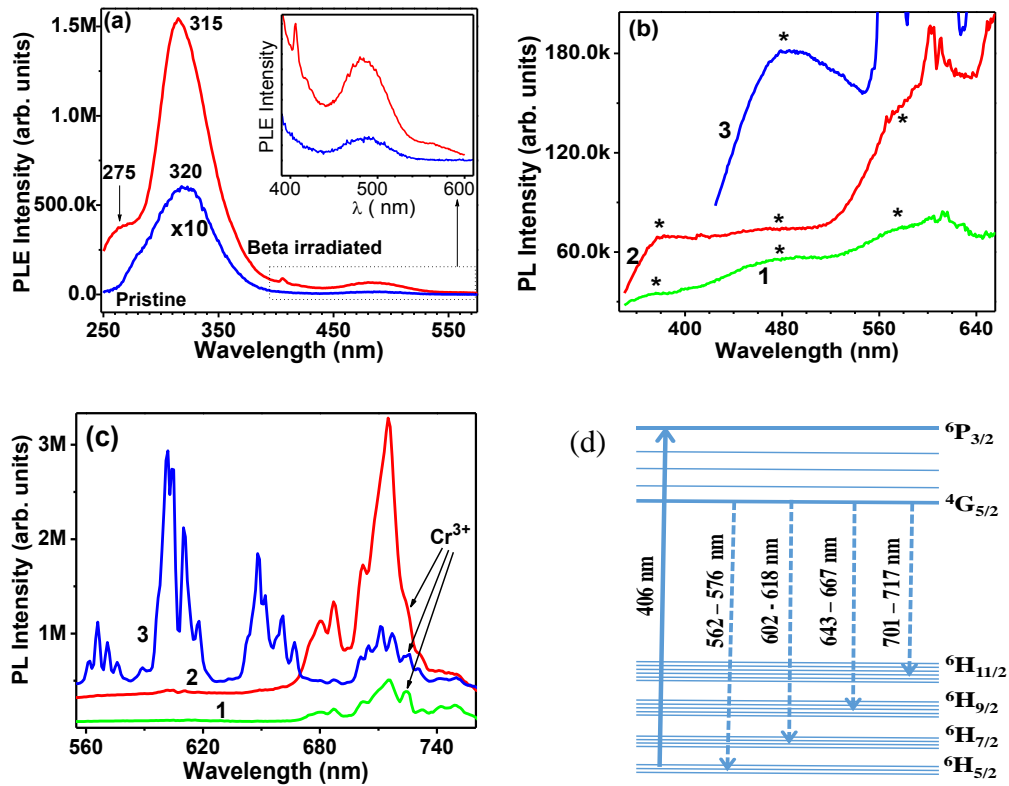

Fig. 5 (a). The PL excitation spectra of pristine and beta irradiated $YAlO_3$ for 715 nm emission. The intensity of pristine sample is multiplied by 10 to bring into the scale. The inset in the figure shows the magnified part of the plot in the wavelength range of 400 - 600 nm. The PL emission spectra of pristine $YAlO_3$ under the excitation of 275 nm (curve 1), and irradiated $YAlO_3$ for the excitation of 320 (curve 2) and 406 nm (curve 3) in the wavelength range of (b) 350 – 650 nm and (c) 550 – 760 nm. (d) The characteristic multiplet emission lines from the intra-band *4f-4f* transitions of $Sm^{3+}$ ions in $YAlO_3$ lattice.

(a) Fig 5a depicts the PLE spectra for the emission wavelength of 715 nm which shows the strong excitation bands at ~ 320 and 315 nm for pristine and irradiated YAP respectively. The intensity of the excitation band is much higher for irradiated crystal as compared to pristine YAP. Further, these spectra show weak excitation bands at around 275 and 480 nm. The irradiated YAP shows a sharp band at about 406 nm which is not prominent in pristine crystal.

(b) The PL emission spectra for different excitation energies are shown in Fig. 5b and 5c in the 350 - 650 nm and 550 - 760 nm wavelength range respectively. A few broad emission bands at about 370, 480 and 580 nm are observed in these crystals as shown in Fig. 5b. The emission at 370 nm can be attributed to the oxygen related $F^+$ centers whereas the emission lines at 480 and 580 nm may be associated to the aggregates of charged oxygen vacancies, specifically by the dimers of $F^+$ and F centers [13,25].

(c) The PL emission spectrum excited at 406 nm reveals several multiplet emission lines (curve no. 3 in Fig. 5c) at 562, 566, 571, 576 nm; 602, 605, 610, 618 nm; 643, 648, 652, 661, 667 nm; and 701, 705, 711, 717 nm. These multiplet emission lines confirm the presence of unintentionally doped $Sm^{3+}$ impurities in the YAP lattice [26]. Note that the $Sm^{3+}$ ions with $^4F_5$ configuration undergoes several intra-shell transitions which emit sharp multiplet lines in the red-orange region at about 566, 605, 648 and 705 nm corresponding to the transition from excited level $4G_{5/2}$ to ground levels $H_{5/2}, 6H_{7/2}, 6H_{9/2}$ and $6H_{11/2}$ respectively, as shown in Fig. 5d. Also, the PL excitation bands at about 320 nm can be attributed to the parity allowed 4f-5d transitions of $Sm^{3+}$ in YAP lattice [26]. The unintentional $Sm^{3+}$ and $Cr^{3+}$ impurities are originated from the starting materials, $Y_2O_3$ and $Al_2O_3$, respectively.

## 3.5. Thermoluminescence studies

The TL glow curve of unirradiated and beta irradiated (with a dose of 1 Gy) YAP single crystals are shown in Fig. 6. The irradiated crystal exhibits a single prominent glow peak at 239 °C for a heating rate of 5 °C/s. On the other hand, the unirradiated crystal shows no TL signature as expected. The appearance of a single glow curve indicates that there is no formation of complex defects when the sample is exposed to beta irradiation. This single glow curve structure is best suited for dosimetric applications because there will be no interference from the neighboring peaks during thermal readout. In order to understand the TL kinetic phenomenon in the YAP crystal, the TL glow curve parameters viz. trap depth, frequency factor and order of kinetics are calculated using variable heating rate method.

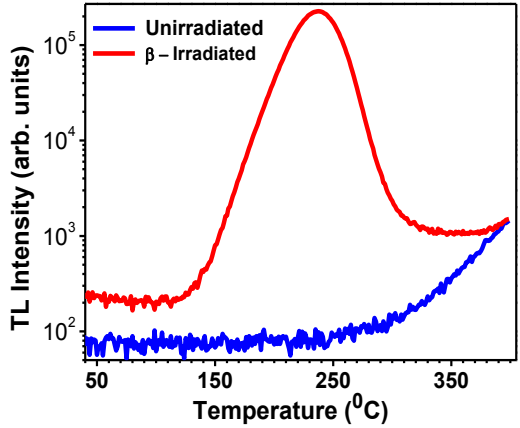

Fig. 6. The TL glow curve of YAlO$_3$ single crystal at heating rate of 5 °C/s.

### 3.5.1 Variable heating rate method

Heating rate is an important parameter which alters the shape and peak position of the glow curve. This property of shift in the glow peak position with heating rate can be exploited for the determination of the kinetic parameters of a TL glow curve. Fig.7a shows the TL glow curve of YAP crystals subjected to a fixed dose of beta radiation of 1 Gy and recorded for different heating rates ranging from 0.5 to 7 °C /s. With increasing heating rate, the TL intensity is found to decrease and also the glow peak temperature ($T_m$) shifts towards higher temperature. In addition, the FWHM of the glow curve also increases with heating rate (Fig.7b). This behavior can be attributed to thermal quenching of TL emission with heating rate which could be due to the finite thickness (1 mm) of the single crystal used for TL measurements [27]. Further, the probability of non-radiative transitions also increases with temperature which causes the reduction of TL intensity. Kitis et al [28] had reported in quartz, LiF, CaF$_2$ phosphors that $T_m$ shifts towards high temperature side and FWHM increases for higher heating rate. Initially, at low heating rate ($\beta_1$), the time spent by the crystal to release all the thermal electrons is relatively high at temperature $T_1$. But for higher heating rate ($\beta_2$), the time spent by the crystal at $T_1$ is reduced so the release of thermal electrons also get reduced and hence, the glow peak shifts to higher temperature $T_2$ [29]. The symmetry factor given by $\mu= \delta/\omega$, where $\omega$ and $\delta$ are the FWHM and high temperature half width of the TL glow curve respectively, is determined and it is found to be 0.44. The numerical solution to the peak shape analysis of glow curve for a pure first order kinetics yields the symmetry factor $\mu$ of $\sim$ 0.42 while it is about 0.52 for second order kinetics [30]. The obtained symmetry factor ($\mu = 0.44$), which is very close to 0.42, suggest that the TL process in YAP crystal obeys first order kinetics. For the case of first order kinetics, the glow peak temperature ($T_m$) and the heating rate ($\beta$) can be related as,

$$\frac{\beta E}{kT_M^2} = s \exp\left(-\frac{E}{kT_M}\right) \qquad \ldots\ldots\ldots (1)$$

where, $T_M$ is the temperature corresponding to maximum TL intensity, E is the trap depth, s is the frequency factor and k is Boltzmann constant.

Fig. 7c shows the plot of $\ln(T_m^2/\beta)$ versus $1/kT_m$ and it would be a straight line according to the equation (1). However, the curve 1 in Fig. 7c displays a large deviation from linearity which might be attributed to the thermal lag between the sample and metallic heating strip due to the finite thickness of the crystal used for TL measurements. Using the simple method suggested by Kitis and Tuyn [30], the peak temperature was corrected for thermal lag using the following equation,

$$T_{mj} = T_{mi} - c \ln\frac{\beta_i}{\beta_j} \qquad \ldots\ldots\ldots (2)$$

where as $T_{mj}$ and $T_{mi}$ are the peak temperature with respect to heating rates of $\beta j$ and $\beta i$ respectively, c is a constant whose value is calculated at low heating rate of 0.5 and 1 °C/s, where thermal lag is assumed to be negligible. Based on equation 2, an effective heating rate ($\beta_e$) is estimated using thermal lag correction. After including the thermal lag correction, the curve 2 in Fig. 7c shows the straight line behavior. Further, the activation energy and frequency factor are calculated using the slope and intercept of the straight line fit (curve 2 of Fig. 7c) and are found to be 1.19 eV and $2.67\times10^{11}$ s$^{-1}$, respectively. Apart from variable heating rate method, we also applied initial rise method to calculate the kinetic parameters, involved in the TL process. The obtained activation energy and frequency factor are found to be 1.42 eV and 2.23 x $10^{12}$ s$^{-1}$ respectively, for heating rate of 5 ºC/s and these values are consistent with the parameters extracted from variable heating rate method.

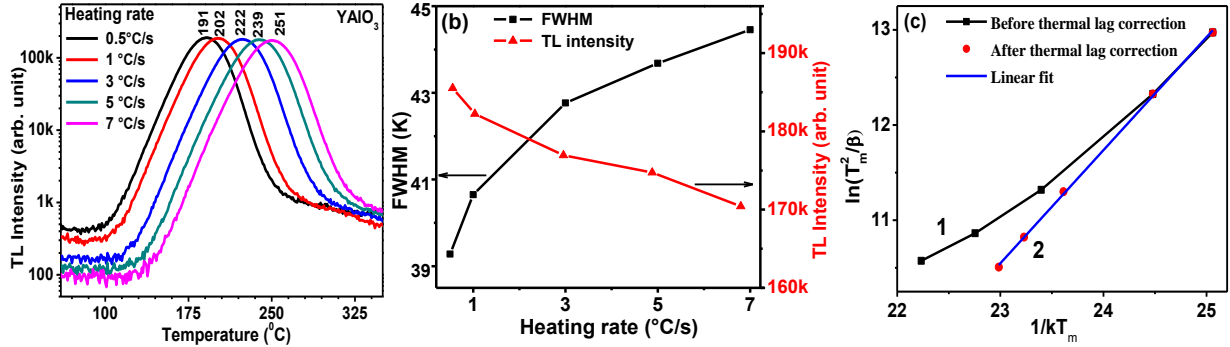

Fig.7. (a) TL glow curves of YAlO$_3$ single crystal recorded at different heating rates (b) Heating rate dependence of FWHM and peak intensity of TL glow curves. (c) The plot of $1/kT_m$ vs $\ln(T_m^2/\beta)$ shows the linear (curve 2) and non-linear curves (curve 1) for with and without accounting thermal lag correction, respectively. The solid lines in Fig. 7b and in curve 1 of Fig. 7c are only guideline to eye.

### 3.5.2 Dose response

Figure 8a displays the TL glow curves of YAP crystals exposed to different beta doses varying from 0.1 to 10 Gy doses and recorded at a heating rate of 5 C/s. A prominent glow curve is obtained at 239 $^0$C for the YAP crystals upto beta dose of 10 Gy. However, the glow curve has a broad shoulder at around 80 $^0$C for 5 Gy dose and it evolves as weak second glow peak for beta dose of 10 Gy. Further, the main glow peak position does not vary with dose thereby indicating that the kinetics is nearly first order. Fig. 8b shows the variation of integrated TL intensity of the glow curve as a function of dose displaying a nearly linear response in the studied dose range of 0.1 – 10 Gy. Moreover, the TL emission is very sensitive for the doses as low as of 100 mGy. Though the YAP crystals have excellent sensitivity, the TL emission could not be carried out at less than 100 mGy doses because of the experimental limitations.

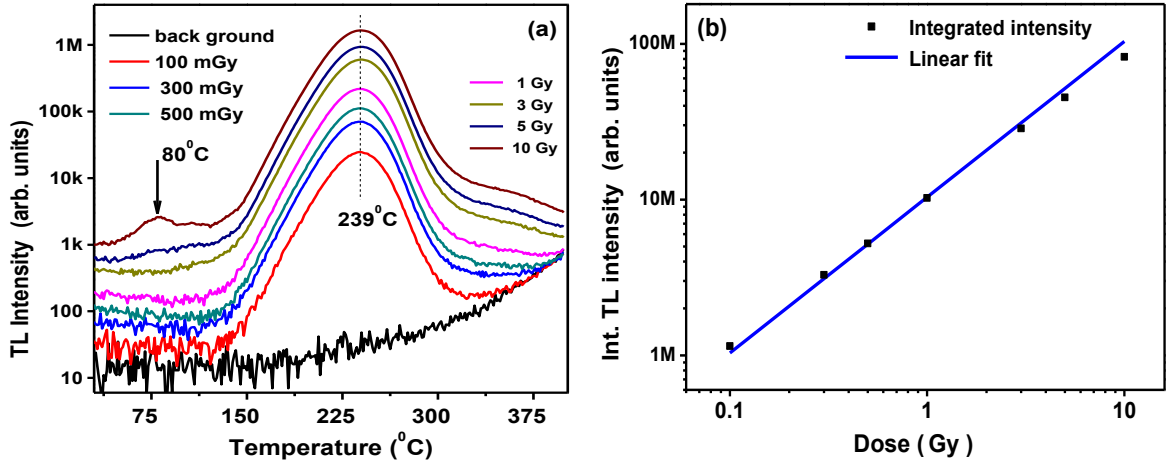

Fig. 8. (a) TL glow curves of YAlO$_3$ at different doses of beta irradiation. (b) The linear response of integrated TL intensity of glow curve as a function of dose in the range of 0.1 to 10 Gy.

### 3.6 Thermally stimulated luminescence spectroscopy

Figure 9 depicts the typical TL emission spectrum of YAP crystal irradiated to a gamma dose of 1 Gy. The spectrum shows intense and broad emission bands at around 566, 604 and 648 nm. As discussed in the PL spectroscopy analysis, it is clear that these emission lines are from the intra-shell *4f-4f* transition of *Sm$^{3+}$* ions in the YAP lattice. Since the multiplet of emission lines overlap with each other, these spectra appear to be broad and asymmetric. Further, there is no TL emission in the wavelength range from 250 to 540 nm which indicates the negligible contribution from intrinsic defects such as oxygen vacancies and antisite defects. Thus, TL emission spectroscopic studies confirm that the TL intensity arises only from the characteristic *4f-4f* intra-shell transitions of *Sm$^{3+}$* impurities in YAP lattice.

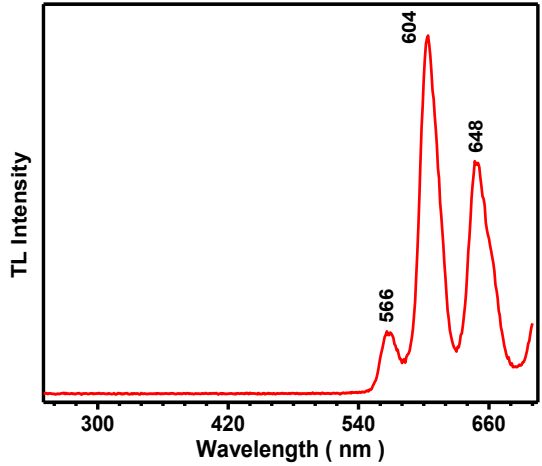

Fig. 9. Thermally stimulated luminescence spectrum of gamma irradiated YAlO$_3$ crystal measured at 200$^0$C.

## 4. Discussion

The TL studies display a single prominent glow curve for YAP crystals subjected to beta irradiation. This behavior dictates that the YAP crystals do have one trap / recombination site in the forbidden bandgap. Further, the PL studies show several emission lines ~ 370, 480 and 580 nm; characteristic emission lines in the wavelength range 560 – 720 nm; and 723 and 725 nm. These bands indicate the presence of oxygen related F/F$^+$ centers and their excitonic emissions; and unintentional impurities such as *Sm$^{3+}$* and *Cr$^{3+}$* in the YAP crystals, respectively. Since the YAP crystals are grown at high temperature, the intrinsic point defects such as oxygen vacancies, antisite defects (Al$_Y$ & Y$_{Al}$) and their defect complexes are unavoidable. These structural defects and the unintentional dopants introduce a lot of intermediate energy levels in the forbidden bandgap. These energy levels trap either electrons or holes that are created during beta irradiation, depending upon the electronic structure of such trap state. In addition, Raman spectroscopic measurements on YAP crystals, before and after irradiation, also confirm the existence of such a large trap states. However, all the trap states may not participate in the thermally simulated luminescence process since some of trap defects are optical active and they undergo bleach effect as observed under 532 nm laser irradiation.

There are a few reports available in literature from experimental and theoretical calculations on the electronic structure and the nature of charge trapping behavior of lanthanide atoms in RE aluminum perovskite (REAlO$_3$) compounds [10,16,18,31]. For example, the trivalent

RE atoms such as *Nd, Dy, Ho, Er, Tm, Sm, Eu* behave as electron traps, while other trivalent atoms such as *Ce, Tb and Pr* act as hole traps in wide bandgap materials [5]. It has been reported by atomistic simulation that the cationic antisite defects are the dominant intrinsic defects over other point defects in YAP [11]. Further, the density functional calculations showed that the *Al* antisite defects ($Al_Y$) act as electron traps with depth of about 0.4 eV below the conduction band edge while the $Y_{Al}$ antisite defects do not produce hole traps [31]. However, the $Y_{Al}$ antisite defects behave as electron trap with paramagnetic $Y_{Al}^{2+}$ structure as evidenced with electron paramagnetic resonance [10]. Further, oxygen anions in YAP crystals trap the holes that are created during irradiation and form as $O^-$ defects which have six different electronic configuration and the trap depth is in the range of 0.024 – 0.5 eV from the valence band [10]. On the other hand, oxygen vacancies act as electron traps in the YAP. Also, the $Al_Y$ & $Y_{Al}$ antisite defects with oxygen vacancy pair act as electron traps in YAP crystals [10].

Based on the experimental observation and the literature knowledge, we can understand the TL emission mechanism on the studied YAP crystals as follows: During the irradiation, the electron – hole pairs are generated. Subsequently, the electrons are trapped by the $Sm^{3+}$ impurity atoms and it becomes $Sm^{2+}$ charge state. Concurrently, the holes are trapped by the $Cr^{3+}$ impurities and $O^-$ defects [10]. When the sample is heated, the trapped holes at $Cr^{3+}$ impurities and intrinsic defects are de-trapped to the valence band and then, these de-trapped holes recombine with the trapped electrons on $Sm^{2+}$ site. Subsequent to the recombination process, the $Sm^{2+}$ ions convert back into $Sm^{3+}$ ions which generate the characteristic red-orange multiplet emission lines due to intra-shell *4f-4f* transitions, as shown in Fig. 5d. Thus, the impurity $Sm^{3+}$ ions in YAP acts as electron trap as well as recombination center. In the present study, the trap depth for the glow peak at 239 $^0$C is measured to be ~ 1.2 eV. Since the trap depth of holes are lower in the range of 0.024 – 0.5 eV [10], the recombination occurs due to hole release in contrast to the commonly known as electron release. Moreover, the weak second glow curve at 80 $^0$C for higher doses ( > 5 Gy) may arise due to the defect complex formation between the intrinsic defects and the *Cr* and *Sm* impurities [16]. Further studies are required to probe the exact origin of hole trap centers that are involved in the TL emission mechanism of YAP crystals.

## 5. Conclusion

The structural and optical properties of YAlO$_3$ (YAP) single crystals grown by optical floating zone technique are studied. The grown crystals have single phase as verified by powder X-ray diffraction and Raman spectroscopy. Further, these intentionally undoped crystals demonstrate excellent thermoluminescence (TL) characteristics with single glow curve at a maximum glow temperature of 239 °C. The crystals also exhibit a nearly linear dose response in the studied range from 0.1 to 10 Gy for beta irradiation. The TL emission mechanism follows first order kinetics and the calculated activation energy is found to be 1.2 eV in YAP crystals. A detailed analysis of TL and optical studies reveal that the unintentional impurity *Sm*$^{3+}$ ions play the major role as electron trap centers as well as recombination centers for TL emission mechanism. The TL and optical studies conclude that a trace level impurities of *Sm*$^{3+}$ and *Cr*$^{3+}$ ions can tune the optical properties significantly. Also, the studied YAP single crystals can be used as a potential candidate for technological applications towards radiation dosimetry.

**Declaration of Competing Interest**

There are no conflicts to declare.

**Acknowledgements**

The authors, M.S & A.S.G, gratefully acknowledge the University Grants Commission, India under UGC-DAE-CSR (CSR-KN/CRS-87/2016-17/1128) project scheme for financial assistance. Also, the authors, M.S and A.S.G, thank Dr. N.V. Chandrasekar, Scientist in-charge, UGC-DAE-CSR, Kalpakkam node for his constant support and encouragement. The author, K.G, thanks Dr. G. Mangamma, Head, NCSS/SND/MSG/IGCAR for her support and encouragement. Also, authors acknowledge Mrs. Sunitha Rajakumari, SND/MSG/IGCAR for her support in crystal polishing.